\documentclass[twoside,twocolumn,9pt]{article}
\usepackage{extsizes}
\usepackage[super,sort&compress,comma]{natbib} 
\usepackage[version=3]{mhchem}
\usepackage[left=1.5cm, right=1.5cm, top=1.785cm, bottom=2.0cm]{geometry}
\usepackage{mathptmx}
\usepackage{sectsty}
\usepackage{graphicx} 
\usepackage[format=plain,justification=justified,singlelinecheck=false,font={stretch=1.125,small,sf},labelfont=bf,labelsep=space]{caption}
\usepackage{float}
\usepackage[english]{babel}
\addto{\captionsenglish}{%
  
}
\usepackage{droidsans}
\usepackage{charter}
\usepackage[T1]{fontenc}
\usepackage[usenames,dvipsnames]{xcolor}
\usepackage{setspace}
\usepackage[compact]{titlesec}
\usepackage{hyperref}
\usepackage{epstopdf}

\begin{document}

\title{\LARGE Corrections to the Fick-Jacobs Description of Diffusion across Pickering Droplet Interfaces}

\author{ Rohit Khandoori$^{1}$, Wim Thielemans$^{1}$, Joost de Graaf$^{2}$ } \date{} 

\twocolumn[ 
\maketitle 
\begin{center}\normalsize 
$^{1}$ Sustainable Materials Lab, Department of Chemical Engineering, KU Leuven, Campus Kulak Kortrijk,\\ Etienne Sabbelaan 53, 8500, Kortrijk, Belgium \vspace{0.3cm} 
\\$^{2}$ Institute for Theoretical Physics, Department of Physics, Faculty of Science, Utrecht University,\\ Princetonplein 5, Utrecht 3584 CC, The Netherlands \vspace{0.3cm} \\Corresponding authors:\\ 
\texttt{wim.thielemans@kuleuven.be}\\ \texttt{j.degraaf@uu.nl} \end{center} \vspace{0.4cm} 
\begin{center} \parbox{0.92\textwidth}{ Particle-laden interfaces give rise to important classes of multiphase materials, including Pickering-Ramsden systems and bicontinuous interfacially jammed emulsion gels (bijels). In these systems, colloidal particles adsorb at fluid-fluid interfaces and form semipermeable layers that provide structural stability and controlled diffusive transport. These properties make them attractive for encapsulation applications and motivate understanding transport across particle-laden interfaces. Diffusion through geometries with varying cross-section, including those found at particle-laden interfaces, is commonly described using the Fick-Jacobs formalism. However, this approach relies on the assumptions of slowly varying geometries and rapid equilibration in the transverse direction. Subsequent studies have developed corrections based on higher-order geometric derivatives, but an elegant closed-form description remains elusive. Here, we compare predictions of the Fick-Jacobs model with fully resolved simulations of 2D Pickering droplets. We find that discrepancies in model predictions increase with interfacial particle crowding, revealing the limits of the Fick-Jacobs approximation. We show that the required correction can be understood through simple geometrical arguments. Our results provide an effective framework for describing transport across particle-laden interfaces and lay the foundation for a more general description of diffusion in complex geometries.\\ } \end{center} ]

\section{Introduction}

Diffusive transport is a fundamental mass-transfer mechanism that governs the exchange of chemical species in a wide range of physical, chemical, and biological systems.\cite{crank1979mathematics,Jacobs1967,cussler2009diffusion} In biological microenvironments such as tissues, biofilms, and intracellular media, which are often structurally heterogeneous, diffusion plays an important role in the transport of nutrients, metabolites, and signaling molecules.\cite{penta2019porosity, karygianni2020biofilm, s2020getting, quan2022water} In chemical and process engineering, diffusion governs the transport of reactants and products through porous catalysts, drives selective species transport across membranes, and mediates mass transfer between coexisting phases in multiphase systems. \cite{van1992diffusional, panda2025diffusion, zid2018modeling} Mass transfer in such systems often occurs across structured interfaces that impose additional constraints on diffusion and lead to deviations from ideal bulk behavior. \cite{meigel2022dispersive} Structured interfaces constitute an important class of transport-regulating structures in natural and engineered systems.\cite{ji2020interfacial,penta2019porosity} Consequently, establishing quantitative relationships between microstructure and diffusive flux has attracted considerable attention across a broad range of disciplines.\cite{zid2018modeling,hunt2017flow}

Particle-laden interfaces are an important subclass of structured interfaces that arise in a variety of multiphase systems. Historically, they have given rise to important classes of soft materials, most notably Pickering emulsions,\cite{de2022review} first reported in 1907,\cite{pickering1907cxcvi} and bicontinuous interfacially jammed emulsion gels (bijels),\cite{clegg2020bijels} proposed in silico in 2005\cite{stratford2005colloidal} and experimentally realized in 2007.\cite{herzig2007bicontinuous} In these systems, colloidal particles adsorb at the boundary between two immiscible fluids and form a semipermeable interfacial layer.\cite{ji2020interfacial} This layer separates the phases while enabling controlled transport between them. Pickering emulsions have been studied for more than a century and are widely used in food, pharmaceutical, and cosmetic formulations for their stabilizing qualities.\cite{de2022review,robin2022pickering} Pickering emulsions can also be used to encapsulate active materials and this application has attracted increasing attention over the past three decades.\cite{velev1996assembly, velev1996assembly2} This growing interest stems from the unique combination of structural stability and tunable transport properties offered by particle-laden interfaces.\cite{yang2017overview} As a result, Pickering emulsions provide a robust platform for the encapsulation and controlled release of diverse functional agents, including probiotic microorganisms for gut health,\cite{haji2022application} microbial pesticides for agricultural applications,\cite{yaakov2018single} and catalysts for biphasic chemical reactions.\cite{rodriguez2020catalysis} Beyond these applications, Pickering-based capsules have also emerged as promising platforms for cell encapsulation in tissue engineering\cite{das2017high} and may offer opportunities for future therapeutic treatments for diabetes, haemophilia, and cancer.\cite{orive2003cell} These capsules protect encapsulated cells while enabling the exchange of nutrients, metabolites, and signalling molecules required for tissue engineering and therapeutic applications. Therefore, rational design and optimization in this emerging field require an understanding of how particle-laden interfaces regulate mass transfer, further motivating the study of diffusion across structured interfaces.

Historically, the transport properties of particle-laden interfaces have been explored in relation to evaporation,\cite{yong2016nanoparticle} gas exchange,\cite{poulichet2015cooling} and solute diffusion.\cite{rosenberg2011diffusion} Berg and Purcell showed that, under steady-state diffusion, a surface containing many well-distributed absorbing pores can exhibit a capture rate approaching that of a perfectly absorbing sphere.\cite{Berg1977} Remarkably, this behavior can occur even when only a small fraction of the surface is accessible to diffusing species. This result provides a useful conceptual framework for the study of diffusive transport across particle-laden interfaces, where mass transfer is mediated by the pores and gaps formed between adsorbed particles. However, for encapsulation purposes there is a need to study systems away from steady state. When considering transient behavior, the literature provides conflicting viewpoints. Several experimental and theoretical studies suggest that particle layers can hinder transport by reducing the exposed interfacial area available for diffusion.\cite{yong2016nanoparticle,dan2012transport} In contrast, other studies indicate that transport rates can remain largely unaffected, even at high particle coverages.\cite{prakash2025evaporation,schroder2019can} This apparent discrepancy has motivated recent efforts to develop unified descriptions of diffusion across particle-laden interfaces. For example, Garbin and co-workers combined experiments and theory to investigate transient diffusion across particle-laden interfaces in Pickering droplets.\cite{liu2023diffusion, van2025scaling, oratis2026diffusion} They developed a Fick-Jacobs description to explain how particle coverage and particle size influence mass transfer, identifying scaling laws that provide a unified interpretation of these apparently conflicting observations.\cite{van2025scaling} However, while the resulting scaling laws are consistent with several experimental observations, the underlying transport equation has not been directly validated against fully resolved experimental concentration fields. Thus, it remains unclear when the coarse-grained Fick-Jacobs model breaks down and when fully resolved particle-scale effects must be explicitly considered.

Examining the assumptions of Garbin and co-workers sheds light on these concerns. Their Fick-Jacobs description relies on rapid equilibration in the direction transverse to the interface and smooth geometric variation.\cite{Jacobs1967, zwanzig1992diffusion, reguera2001kinetic, kalinay2006corrections} These assumptions effectively replace the complex pore network formed by adsorbed particles with a homogenized, quasi-one-dimensional transport description. The extent to which this approximation remains valid for densely packed particle layers containing sharp constrictions and bottlenecks remains largely unexplored. Here, we therefore revisit diffusion through particle-laden Pickering droplet interfaces using two-dimensional simulations that explicitly resolve particles adsorbed at the droplet interface. We compare the resulting diffusion profiles with predictions from the corresponding quasi-one-dimensional Fick-Jacobs model. Our comparison reveals systematic deviations between the reduced model and simulations as interfacial particle crowding increases. To
address these deviations, we incorporate a curvature-corrected description of the Pickering droplet geometry and introduce a simple moment-based correction to the quasi-one-dimensional
model. The resulting framework reproduces simulation results while incorporating particle-induced tangential diffusion, a mechanism neglected in the original quasi-one-dimensional description.

\section{Model and Methods}

We considered diffusive transport across the particle-laden interface of a two-dimensional Pickering droplet geometry. The system consists of a circular droplet of radius $R$ with $N$ colloidal particles of radius $a$ adsorbed at the interface, as illustrated in Fig.~\ref{fig:1}(c--f). Particles are modeled as disks whose centers lie on the droplet circumference at $r=R$. The particle size is characterized by the dimensionless ratio $r_p = \frac{a}{R}$. The particle coverage is described by the interfacial coverage parameter $\phi_0$, defined as the fraction of the interfacial circumference at $r=R$ blocked by adsorbed particles, 

\begin{equation}
\phi_0 = \frac{N(2a)}{2\pi R} = \frac{Na}{\pi R}.
\label{eq:phi_0} 
\end{equation} 

For instance, a value of $\phi_0=0.80$ with $r_p=0.05$ corresponds to $N=50.3$ particles according to Eq.~\eqref{eq:phi_0}. Since the number of particles must be an integer, the corresponding configuration contains $N=50$ particles, resulting in $\phi_0=0.796$. Thus, the values of $\phi_0$ considered in this study were calculated from the integer particle configurations for each combination of $N$ and $r_p$.

To understand the diffusive process in this system, we compare a quasi-one-dimensional Fick--Jacobs model with fully resolved two-dimensional simulations.\cite{liu2023diffusion,van2025scaling} By comparing the two approaches, we assess the validity of the quasi-one-dimensional model and identify corrections arising from finite particle size, interfacial curvature, and particle-scale transport effects.

\subsection{Quasi-one-dimensional Fick--Jacobs model}

\begin{figure}[t]
    \centering
    \includegraphics[width=0.40\textwidth]{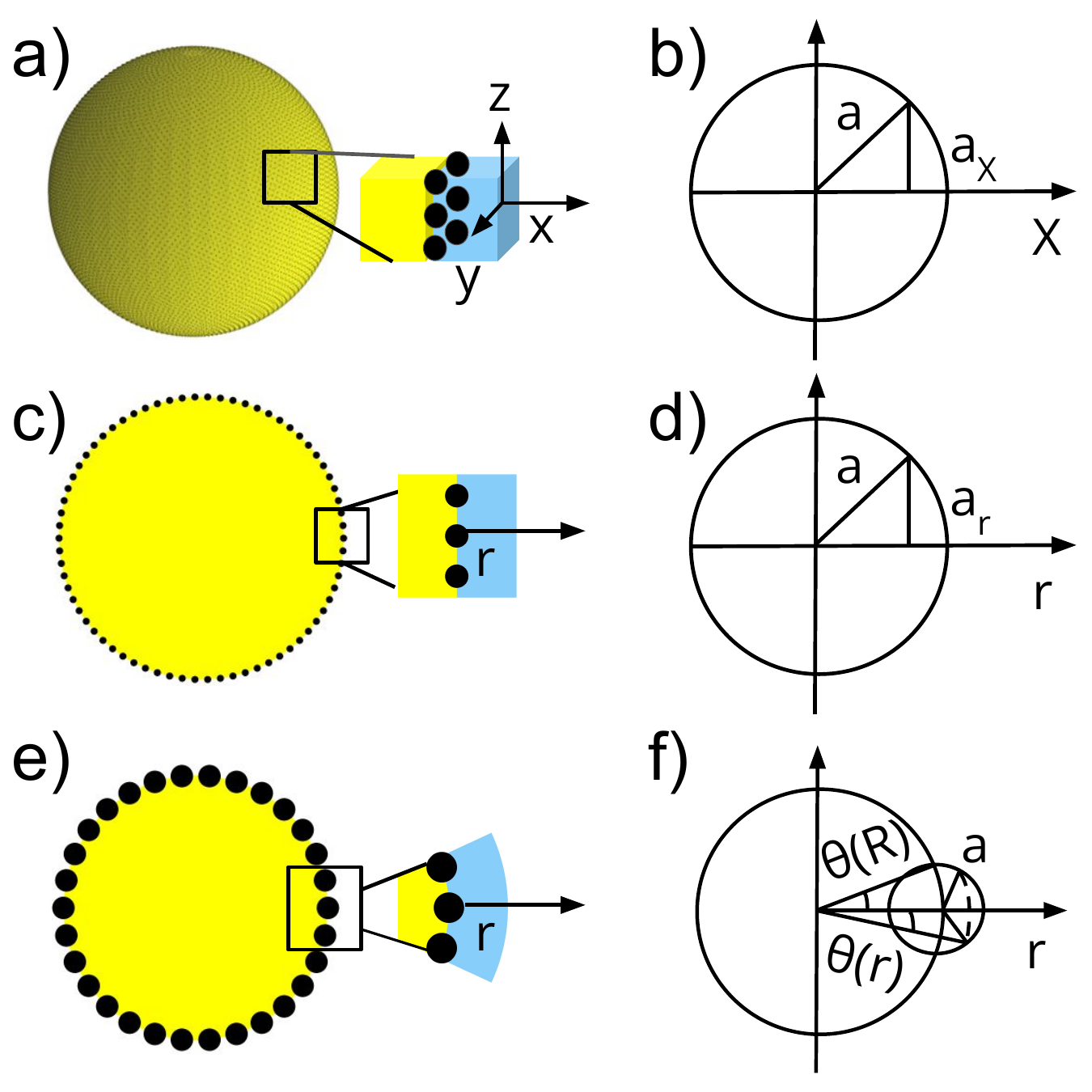}
    \caption{ Geometric interpretation of the quasi-one-dimensional description of diffusion across particle-laden interfaces. (a) Three-dimensional Pickering droplet with spherical particles adsorbed at the fluid--fluid interface. A local enlargement illustrates the particle-laden interfacial region through which diffusion occurs. (b) Planar $x$--$z$ cross-section underlying the original Fick--Jacobs formulation, showing the projected particle radius $a_x$ used to define the local obstruction profile. (c) Two-dimensional circular droplet with particles adsorbed along the interface. (d) Radial cross-section of the two-dimensional geometry, where the particle intercepts a chord of length $2a_r$ at radial position $r$. (e) Particle-laden circular interface illustrating the finite particle size relative to the droplet radius. (f) Geometric quantities used to construct the radial obstruction profile in the curved geometry, including the droplet radius $R$, particle radius $a$, radial coordinate $r$, and angular extent $\theta$ associated with the particle obstruction. }
    \label{fig:1}
\end{figure}

We begin by revisiting the quasi-one-dimensional description proposed by Garbin and co-workers\cite{liu2023diffusion} for diffusion across particle-laden Pickering droplets. The geometric interpretation of this model is illustrated in Fig.~\ref{fig:1}(a,b). Their formulation considers a three-dimensional droplet with spherical particles adsorbed at the interface as in Fig.~\ref{fig:1}(a). In the limit of small particles, $a/R\ll1$, the interface can be approximated locally as a flat interface and diffusion is assumed to occur in the direction normal to the interface. 

Under these assumptions, the discrete particle layer is replaced by a continuous obstruction profile that accounts for the reduction in the interfacial area available for diffusion. The obstruction profile is obtained by projecting the cross-sectional area of a spherical particle onto the diffusion direction, as illustrated in Fig.~\ref{fig:1}(b). This projection defines a local blocked fraction $\phi(x)$, while the complementary fraction $A(x)=1-\phi(x)$ represents the area available for transport. For a spherical particle of radius $a$, the resulting obstruction profile is 

\begin{equation} 
\phi(x)= \begin{cases} \displaystyle \phi_0\frac{a^2-x^2}{a^2}, & -a\le x\le a, \\[8pt] 0, & |x|>a , 
\end{cases} 
\label{eq:phi_planar} 
\end{equation} 

where $\phi_0$ denotes the maximum interfacial coverage and therefore corresponds to the maximum value of the obstruction profile at the interface. Within this coarse-grained description, diffusion through the particle-laden interfacial region is governed by the Fick--Jacobs equation 

\begin{equation} (1-\phi(x))\frac{\partial C}{\partial t} = \frac{\partial}{\partial x} \left[ D(1-\phi(x)) \frac{\partial C}{\partial x} \right], \label{eq:FJ_planar} \end{equation}

where $C(x,t)$ denotes the solute concentration and $D$ is the molecular diffusivity. Equation~\eqref{eq:FJ_planar} retains the form of Fickian diffusion, with an effective diffusion coefficient $D_A$ within the interfacial region. From its definition, $D_A=D\,A(x)$, indicating that $D_A$ is determined by the molecular diffusivity $D$ scaled by the local area fraction $A$ available for transport. The Fick--Jacobs formulation assumes rapid equilibration in directions transverse to transport,\cite{Jacobs1967, zwanzig1992diffusion} allowing the inherently three-dimensional diffusion problem to be reduced to an effective one-dimensional description.

In the present work, we consider a two-dimensional circular droplet with finite-sized particles adsorbed along the interface, as illustrated in Fig.~\ref{fig:1}(c--f). In contrast to the planar geometry considered in the original formulation,\cite{liu2023diffusion} diffusion in a circular droplet is inherently  radial, making the Cartesian coordinate system less suitable. Furthermore, interfacial curvature cannot be neglected because of finite particle-to-droplet size ratios. For direct comparison with fully resolved two-dimensional simulations, we adopted the Fick--Jacobs description in cylindrical coordinates and assume azimuthal symmetry, such that transport is described by the radial concentration field $C(r,t)$.\cite{oratis2026diffusion} The governing equation becomes

\begin{equation} 
\frac{\partial C}{\partial t} = \frac{1}{r[1-\phi(r)]} \frac{\partial}{\partial r} \left[ rD(1-\phi(r)) \frac{\partial C}{\partial r} \right], \label{eq:FJ_radial} 
\end{equation}

where $\phi(r)$ denotes the effective radial obstruction profile generated by particles adsorbed at the droplet interface. The specific form of $\phi(r)$ determines how particle geometry and interfacial curvature are incorporated into the quasi-one-dimensional description and is discussed in the following section.

\subsection{Geometric correction for finite particle size}

 Following the recent extension of the Fick--Jacobs framework to cylindrical coordinates,\cite{oratis2026diffusion} the obstruction profile is obtained by projecting a spherical particle onto the radial coordinate, yielding

\begin{equation} 
\phi_{\mathrm{sphere}}(r)= \begin{cases} \displaystyle \phi_0\frac{a^2-(r-R)^2}{a^2}, & |r-R|\le a, \\[8pt] 0, & |r-R|>a . 
\end{cases} 
\label{eq:phi_sphere} 
\end{equation} 

Although Eq.~\eqref{eq:phi_sphere} provides a convenient radial representation of the particle obstruction, it is based on the projection of a three-dimensional spherical particle. In the present work, we considered a two-dimensional droplet geometry in Fig.~\ref{fig:1}(c), for which particles are represented by circles. In the limit $a/R\ll1$, the particle obstruction is more naturally described by the chord length intercepted by a circular particle cross-section, as illustrated in Fig.~\ref{fig:1}(d). This leads to the chord-based profile

\begin{equation} 
\phi_{\mathrm{chord}}(r)= \begin{cases} \displaystyle \phi_0 \frac{\sqrt{a^2-(r-R)^2}}{a}, & |r-R|\le a, \\[10pt] 0, & |r-R|>a . 
\end{cases} 
\label{eq:phi_chord} 
\end{equation} 

Equation~\eqref{eq:phi_chord} accounts for the circular particle geometry through the intercepted chord length. While this description is appropriate for small particle-to-droplet size ratios, interfacial curvature becomes increasingly important as $a/R$ increases as in Fig.~\ref{fig:1}(e). In this regime, the available diffusion pathway is more accurately described by the intercepted arc length along the circular interface, as illustrated in Fig.~\ref{fig:1}(f). To account for this effect, we compute the angular intersection between a particle of radius $a$ and a droplet of radius $R$. The corresponding angle $\theta(r)$ satisfies 

\begin{equation} 
\cos\theta(r) = \frac{R^2-a^2+r^2}{2Rr}. 
\label{eq:theta_arc}
\end{equation} 

The curvature-corrected obstruction profile is then defined as 

\begin{equation} \phi_{\mathrm{arc}}(r)= \begin{cases} \displaystyle \phi_0 \frac{\theta(r)}{\theta(R)}, & |r-R|\le a, \\[10pt] 0, & |r-R|>a , \end{cases} \label{eq:phi_arc} \end{equation}

The normalization by $\theta(R)$ ensures that $\phi_{\mathrm{arc}}(R)=\phi_0$ at the droplet interface. In the limit of small particles,  Eq.~\eqref{eq:phi_arc} approaches the chord-based approximation in Eq.~\eqref{eq:phi_chord}. By explicitly incorporating finite-size particles and curvature effects, Eq.~\eqref{eq:phi_arc} provides the most geometrically consistent representation for the geometries considered here and is therefore used throughout this work.

\subsection{Fully resolved two-dimensional simulations}

To test the validity of the reduced model, we solve the full two-dimensional diffusion equation

\begin{equation}
    \frac{\partial C}{\partial t}
    =
    D\nabla^2 C
\label{eq:2D_diffusion}
\end{equation}

in a circular droplet geometry with particles explicitly resolved at the interface. Figure~\ref{fig:2}(a) illustrates the simulation geometry, where slight deviations from perfectly circular particle boundaries reflect the numerical grid representation. The particles are treated as impenetrable obstacles, and no-flux boundary conditions are imposed on their surfaces,

\begin{equation}
    \mathbf{n}\cdot\nabla C = 0,
\label{eq:no_flux_particles}
\end{equation}

where $\mathbf{n}$ is the normal to the particle boundary. 

The diffusion field is initialized using a Heaviside concentration profile,

\begin{equation}
C(\mathbf{x},0)=
\begin{cases}
C_0, & r<R,\\
0, & r>R,
\end{cases}
\label{eq:initial_condition}
\end{equation}

where $r=\sqrt{x^2+y^2}$. The diffusion equation is solved in a square computational domain extending from $-3R$ to $3R$ in both spatial directions. Dirichlet boundary conditions, $C=0$, are imposed at the outer boundaries of the domain ($x=\pm3R$ and $y=\pm3R$). The concentration is reported in dimensionless form as $c=C/C_0$, and time is nondimensionalized as

\begin{equation}
    T=\frac{Dt}{R^2}.
\end{equation}

The two-dimensional diffusion equation is solved numerically on a Cartesian grid using a finite-difference scheme. The grid resolution was chosen such that further refinement produced negligible changes in the results. The resulting concentration field $C(x,y,t)$ is azimuthally averaged to obtain a radial concentration profile,

\begin{equation}
    \langle C(r,t)\rangle_{\theta}
    =
    \frac{1}{2\pi}
    \int_0^{2\pi}
    C(r,\theta,t)\,d\theta .
\label{eq:azimuthal_average}
\end{equation}

which can be compared directly with the solution of the quasi-one-dimensional model in Eq.~\eqref{eq:FJ_radial}, which is solved numerically using a finite-volume method. For consistency, both the quasi-one-dimensional and fully resolved models employ the Heaviside initial condition in Eq.~\eqref{eq:initial_condition} and Dirichlet boundary conditions at the far boundaries of the computational domain.

\section{Results and Discussion}

We now examine the validity of the quasi-one-dimensional Fick--Jacobs description for diffusion across particle-laden Pickering droplet interfaces. We begin by investigating how different geometric representations of the particle obstruction influence the reduced model. The resulting obstruction profiles are then used to compare the quasi-one-dimensional model predictions with the simulations. Finally, the discrepancies between the two descriptions are analyzed to identify the transport mechanisms absent from the reduced model and to develop a corresponding correction.

\subsection{Effect of obstruction geometry on the quasi-one-dimensional model}

\begin{figure*}[t]
    \centering
    \includegraphics[width=1.0\textwidth]{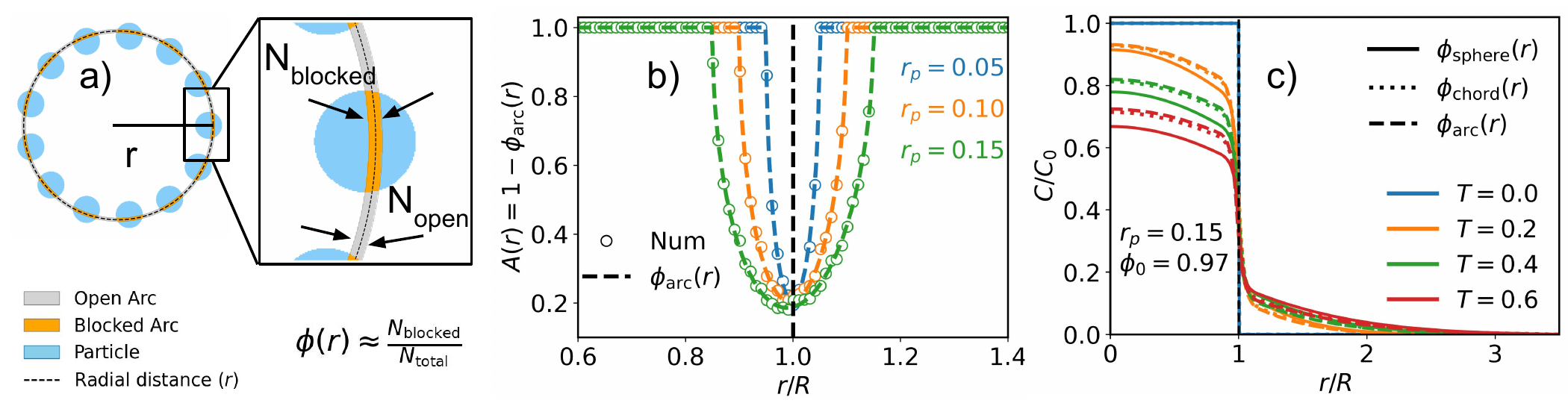}
    \caption{ \textbf{Effect of obstruction geometry on the quasi-one-dimensional diffusion model.} (a) Illustration of the numerical particle configuration used to calculate $\phi_{\mathrm{arc}}(r)$. (b) Available area fraction $A(r)=1-\phi_{\mathrm{arc}}(r)$ for a fixed packing fraction $\phi_0=0.80$ and particle-size ratios $r_p=0.05$, $0.10$, and $0.15$. Open symbols denote the available area fraction measured directly from the numerical particle geometry, while solid lines show the curvature-corrected arc model. (c) Concentration profiles calculated by the quasi-one-dimensional model for $r_p=0.15$ and $\phi_0=0.97$ using the obstruction profiles $\phi_{\mathrm{sphere}}(r)$, $\phi_{\mathrm{chord}}(r)$, and $\phi_{\mathrm{arc}}(r)$. Colors indicate the dimensionless time $T=Dt/R^2$.}
    \label{fig:2}
\end{figure*}

The obstruction profile connects the interfacial particle geometry to the quasi-one-dimensional diffusion model. We first validate the curvature-corrected profile $\phi_{\mathrm{arc}}(r)$ and then evaluate the influence of the obstruction profile on diffusive transport. Figure~\ref{fig:2}(a) illustrates the numerical procedure used to calculate the obstruction profile. The panel shows a representative particle configuration for $r_p=0.15$ having $\phi_0=0.60$ and a magnified view of the discretized particle geometry used in the simulations. At a radial distance $r$, the obstruction is determined by counting the number of grid points lying within particles, $N_{\mathrm{blocked}}$, along a circular ring and normalizing by the total number of grid points on the ring, $N_{\mathrm{total}}$. The resulting quantity defines the numerical obstruction profile $\phi_{\mathrm{arc}}(r)=N_{\mathrm{blocked}}/N_{\mathrm{total}}$, which represents the fraction of blocked points at radial position $r$.

Figure~\ref{fig:2}(b) shows the available area fraction \(A(r)=1-\phi_{\mathrm{arc}}(r)\) for a fixed packing fraction $\phi_0=0.80$ and particle-size ratios $r_p=0.05$, $0.10$, and $0.15$. The solid lines correspond to the analytical prediction obtained from Eq.~\eqref{eq:phi_arc}, while the symbols denote values measured directly from the numerical particle geometry. Excellent agreement is observed over the entire range of particle sizes considered, confirming that Eq.~\eqref{eq:phi_arc} accurately captures the geometric obstruction created by particles adsorbed at the interface. For the smallest particle size, $r_p=0.05$, the available area fraction is nearly symmetric about the interface, indicating that curvature effects are weak and $\phi_{\mathrm{arc}}(r)\approx\phi_{\mathrm{chord}}(r)$. As the particle size increases, the available area fraction departs from the nearly symmetric profile observed for small particles and develops a pronounced skewness toward the droplet interior. This curvature-induced asymmetry is not captured by the chord-based approximation, which assumes a symmetric obstruction profile about the interface. Consequently, the discrepancy between the chord- and arc-based descriptions increases with increasing particle-to-droplet size ratio.

The influence of the obstruction geometry on the diffusive transport is illustrated in Fig.~\ref{fig:2}(c). The concentration profiles are obtained by solving Eq.~\eqref{eq:FJ_radial} using the obstruction profiles $\phi_{\mathrm{sphere}}(r)$, $\phi_{\mathrm{chord}}(r)$, and $\phi_{\mathrm{arc}}(r)$. Since diffusive transport is affected by $D[1-\phi(r)]$, different geometric representations of the particle layer lead directly to different diffusion predictions. The spherical projection, $\phi_{\mathrm{sphere}}(r)$, predicts the fastest diffusion because it provides the largest available area for transport. In contrast, the chord- and arc-based descriptions, $\phi_{\mathrm{chord}}(r)$ and $\phi_{\mathrm{arc}}(r)$, produce a stronger effective obstruction and therefore slower diffusion. For the particle size shown in Fig.~\ref{fig:2}(b), the chord- and arc-based predictions remain relatively close, although the arc profile yields a slightly slower transport because of the effect of interfacial curvature.

Together, Figs.~\ref{fig:2}(b) and (c) demonstrate that the geometric representation of the particle layer has a direct influence on the predicted diffusion dynamics. While the differences between $\phi_{\mathrm{chord}}(r)$ and $\phi_{\mathrm{arc}}(r)$ remain modest for small particles, they arise from the finite curvature of the droplet interface. As the particle size becomes comparable to the droplet radius, these curvature-induced differences are expected to become increasingly important. The arc-based obstruction profile provides the most geometrically consistent description of finite-sized particles adsorbed on a curved interface. We therefore adopt $\phi_{\mathrm{arc}}(r)$ in the subsequent comparisons with fully resolved two dimensional simulations.

\subsection{Comparison between two-dimensional simulations and the arc-corrected quasi-one-dimensional model}

\begin{figure*}[t]
    \centering
    \includegraphics[width=0.9\textwidth]{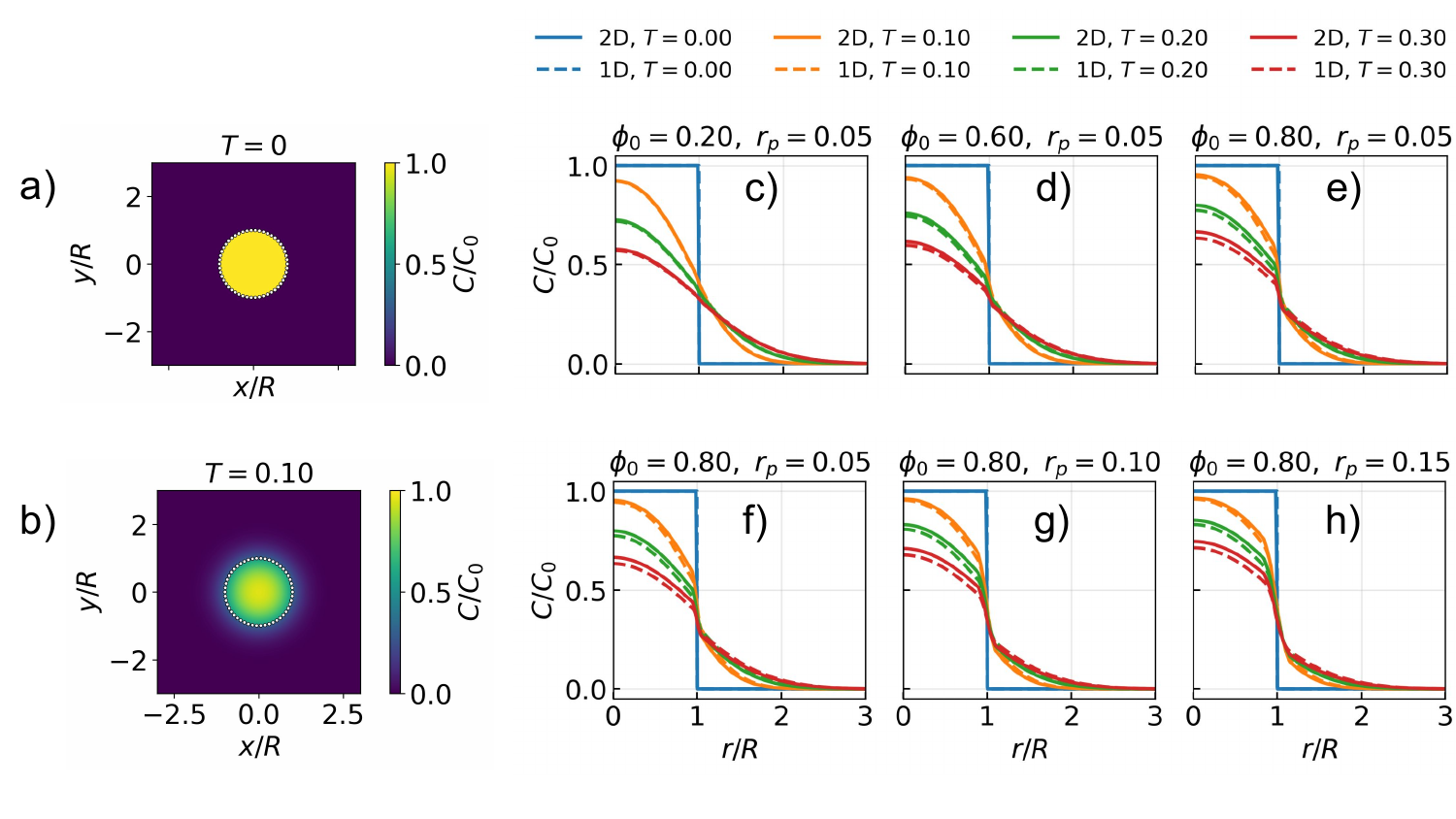}
    \caption{
    \textbf{Comparison between fully resolved two-dimensional simulations and the quasi-one-dimensional model.}
    (a,b) Dimensionless concentration field $C/C_0$ obtained from the two-dimensional simulations at reduced times $T=0$ and $T=0.10$, respectively, for $\phi_0=0.80$ and $r_p=0.05$. The dotted circle indicates the droplet interface.
    (c--e) Comparison between the azimuthally averaged two-dimensional concentration profiles and the quasi-one-dimensional model for fixed particle size $r_{p}=0.05$ and increasing interfacial packing fraction $\phi_0$ .
    (f--h) Comparison for fixed interfacial packing fraction $\phi_0=0.80$ and increasing particle size $r_p$.
    In panels (c--h), solid lines correspond to the two-dimensional simulations, while dashed lines represent the quasi-one-dimensional model. The times at which the profiles are evaluated are indicated in the legend.
    }
    \label{fig:3}
\end{figure*}

Let us now compare the results of the quasi-one-dimensional model with two-dimensional simulations. Figure~\ref{fig:3}(a,b) shows the concentration field obtained from the two-dimensional simulations at $T=0$ and $T=0.10$, respectively, illustrating diffusion across the particle-laden interface. The concentration fields are subsequently azimuthally averaged to obtain radial concentration profiles, which are compared with the quasi-one-dimensional model in Figs.~\ref{fig:3}(c--h).

First, we consider the effect of increasing interfacial packing fraction $\phi_0$ on diffusive transport at fixed particle size $r_p=0.05$, as shown in Figs.~\ref{fig:3}(c--e). For low packing fractions, where the particles remain relatively sparse at the interface, the agreement between the two descriptions is good and the reduced model accurately reproduces the simulated concentration profiles. As the packing fraction increases, the discrepancy between the two descriptions becomes progressively larger. Although the quasi-one-dimensional model appears to capture the overall shape of the concentration profile, it also appears to predict faster diffusion than observed in the simulations. The increasing disagreement at high packing fractions suggests that the homogenized obstruction profile becomes less accurate as the particle layer becomes increasingly crowded.

To determine whether the same behavior occurs with particle size, Figs.~\ref{fig:3}(f--h) show the corresponding comparison for a fixed packing fraction $\phi_0=0.80$ and increasing particle-size ratio $r_p$. Increasing the particle size slows down the diffusion process in both the quasi-one-dimensional model and the fully resolved simulations, leading to broader concentration gradients across the interface and slower relaxation dynamics. However, unlike the packing-fraction dependence, the discrepancy between the two descriptions remains qualitatively similar across the range of particle sizes considered. This observation suggests that the influence of particle size is already captured reasonably well through the obstruction profile $\phi_{\mathrm{arc}}(r)$. Consequently, particle size primarily affects the diffusion timescale rather than the validity of the quasi-one-dimensional model itself.

Our comparison indicates that the dominant source of error is associated with increasing particle crowding rather than particle size. Since finite-size and curvature effects are already incorporated through $\phi_{\mathrm{arc}}(r)$, the remaining discrepancies likely originate from transport mechanisms that are absent from the quasi-one-dimensional description. This indicates diffusive transport can occur through tangential diffusion around individual particles, giving rise to tortuous diffusion pathways that are not represented in the reduced model.

\subsection{Transport correction to the quasi-one-dimensional model} 

\begin{figure}[t]
    \centering
    \includegraphics[width=0.5\textwidth]{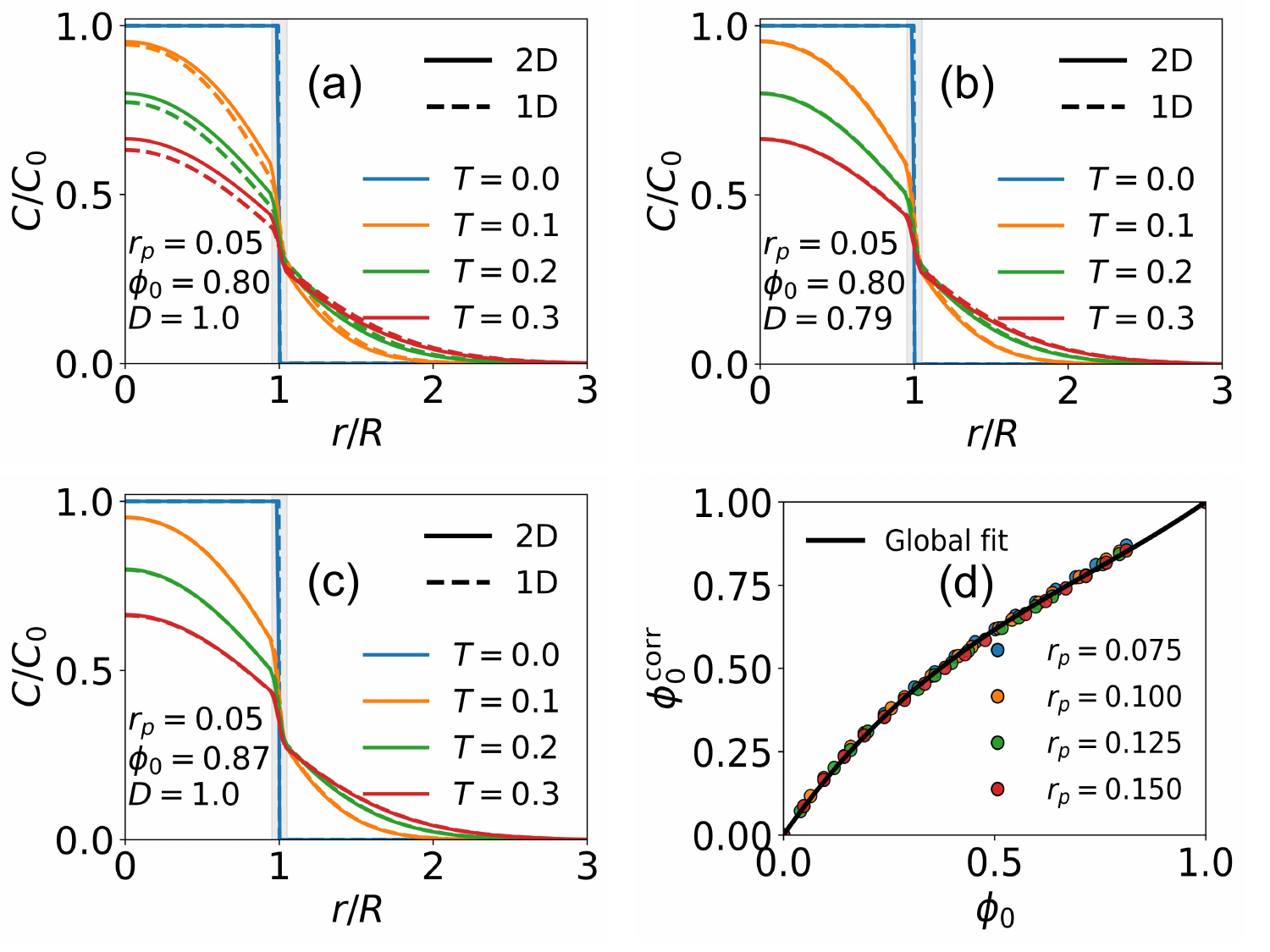}
    \caption{ Transport correction to the arc-corrected quasi-one-dimensional model. (a) Comparison between concentration profiles from fully resolved two-dimensional simulations and the quasi-one-dimensional model for $r_p=0.05$ and $\phi_0=0.80$ having $D=1.0$. 
    (b) Corresponding comparison after fitting the D at interface while keeping $\phi_0$ fixed. (c) Corresponding comparison obtained after fitting $\phi_0$ while keeping $D=1.0$. (d) Corrected packing fraction $\phi_{\mathrm{corr}}$ as a function of the different packing fraction $\phi_0$ for different particle radii. The solid line denotes the global fit given by Eq.~\eqref{eq:phi_eff}. }
    \label{fig:4}
\end{figure}

The observation that diffusion appears accelerated while qualitative trends remain similar suggests that the quasi-1D model can be aligned with 2D simulation data by modifying $D_A$ through $D$ or $A(r)$. Figure~\ref{fig:4}(a) contrasts the regular arc-based prediction with the corresponding simulation results. First, we examined whether this observed discrepancy arises from an incorrect diffusion timescale and can be explained by adjusting $D$ at the interface. Accordingly, the quasi-one-dimensional model of Fig.~\ref{fig:4}(a) remained unchanged, and only $D$ at the interface was treated as a fitting parameter. The interfacial diffusivity $D$ was iteratively adjusted using the Nelder-Mead optimization algorithm to minimize the least-squares error defined in Eq.~\eqref{eq:Fit}, yielding the best-fit profiles shown in Fig.~4(b).

\begin{equation} E(D)=\sum_t \frac{1}{N_r}\sum_r \left[C_{\mathrm{1D}}(r,T)-C_{\mathrm{2D}}(r,T)\right]^2, 
\label{eq:Fit}
\end{equation} 

where $C_{\mathrm{1D}}(r,T)$ and $C_{\mathrm{2D}}(r,T)$ are the quasi-1D and radially averaged 2D concentration profiles, respectively, $N_r$ is the number of radial grid points, and the summation is performed over all recorded times. However, the resulting profiles do not reproduce the two-dimensional diffusion dynamics consistently. Although the overall relaxation becomes slower, the fitted profiles fail to follow the shape and temporal evolution observed in the fully resolved simulations. This indicates that the discrepancy cannot be attributed to a simple rescaling of the diffusivity at the interface.

We next fixed $D$ and repeated the same fitting procedure, treating $\phi_0$ as fitting parameter and obtaining $\phi_0^{\mathrm{corr}}$. As shown in Fig.~\ref{fig:4}(c), this procedure accurately reproduces the concentration profiles when compared with the simulations. We then repeated the same procedure for different particle sizes and packing fractions, obtaining the values of $\phi_0^{\mathrm{corr}}$ shown in Fig.~\ref{fig:4}(d). Remarkably, all data considered in this study for particle radii up to $r_p=0.15$ collapse onto a single curve, and the relationship between $\phi_0$ and $\phi_0^{\mathrm{corr}}$ is described by

\begin{equation}
\phi_0^{\mathrm{corr}} = \phi_0 + a\,\phi_0(1-\phi_0) - b\,\phi_0^2(1-\phi_0), \label{eq:phi_eff} 
\end{equation} 

with fitted coefficients 

\begin{equation} 
a=\frac{\pi}{4}, \qquad b=\frac{2}{3}.  \label{eq:coefficients} \end{equation} 

The geometric origin of these coefficients is discussed later in this section. A least-squares fit of Eq.~\eqref{eq:phi_eff} gives $a=0.804$ and $b=0.665$, differing by only $2.45\%$ and $0.105\%$ from the proposed geometric coefficients. Accordingly, the geometric coefficients used in Fig.~\ref{fig:4}(d) provide an almost identical fit to the data ($R^2=0.99903$). The weak dependence of the fitted coefficients on particle size indicates that the correction is controlled primarily by the packing fraction. To interpret this result, we substitute Eq.~\eqref{eq:phi_eff} into $D_A$

\begin{equation} 
D_A(r) = D \left[ 1 - \phi_0^{\mathrm{corr}} \frac{\theta(r)}{\theta(R)} \right]. \label{eq:DAcorr} 
\end{equation} 

Using Eq.~\eqref{eq:phi_eff}, this becomes \begin{equation} D_A(r) = D \left[ 1 - \phi_0 \frac{\theta(r)}{\theta(R)} \right] - D \left[ a\phi_0(1-\phi_0) - b\phi_0^2(1-\phi_0) \right] \frac{\theta(r)}{\theta(R)}. \label{eq:DAsplit} \end{equation} 

The first term in Eq.~\eqref{eq:DAsplit} corresponds to the original radial Fick--Jacobs description based on the arc-corrected obstruction profile, while the second term represents the additional correction identified through the fitting procedure. It reflects transport mechanisms that are absent from the purely radial formulation. In the fully resolved simulations, solute molecules can diffuse around individual particles through tangential pathways along the interface before continuing their radial transport. These pathways delay mass transfer relative to the idealized radial description assumed by the Fick--Jacobs model. The correction term in Eq.~\eqref{eq:DAsplit} can therefore be interpreted as a phenomenological representation of these unresolved transport processes.

To understand the origin of the coefficients obtained in our corrections Eq.~\eqref{eq:coefficients}, we  consider the development of the Ficks-Jacob equation and the assumptions underlying its derivations. In \textit{Diffusion Processes},\cite{Jacobs1967} M.~H.~Jacobs revisited Fick's (1855) experiments on diffusion through vessels of varying cross-sectional area and developed a heuristic argument based on flux balance to derive an effective one-dimensional transport equation. In this formulation, the confinement geometry enters through the local accessible cross section $A(x)$, such that the diffusive transport is governed by the description $D_0A(x)$. The derivation relies on the assumption of rapid transverse equilibration, implying that the concentration remains locally uniform across each cross section and depends only on the longitudinal coordinate. Later, Zwanzig (1992)\cite{zwanzig1992diffusion} derived the Fick--Jacobs equation systematically from the Smoluchowski equation and analyzed deviations from local transverse equilibrium. Writing the concentration as 

\begin{equation} 
c(x,y,t)=C(x,t)p(y|x)+\delta c(x,y,t), 
\end{equation} 

where $p(y|x)$ is the local equilibrium probability distribution across the channel cross section at a given longitudinal position $x$, he showed that the leading correction can be incorporated through a position-dependent diffusivity. For a two-dimensional channel of half-width $w(x)$, corresponding to an accessible cross section $A(x)=2w(x)$, the effective diffusivity becomes

\begin{equation} 
D(x)\simeq\frac{D_0}{1+\gamma w'(x)^2}, 
\end{equation} where 

\begin{equation} 
\gamma=\frac{1}{w(x)^2}\int dy\, y^2 p(y|x).
\label{eq:moment}
\end{equation}

The resulting transport is therefore governed by $D(x)A(x)$ rather than $D_0A(x)$. This first-order correction extends the validity of the Fick--Jacobs approximation beyond strict local equilibria but remains restricted to slowly varying channels, $|w'(x)|\ll1$. Building on Zwanzig's analysis, Reguera and Rub\'i (2001) \cite{reguera2001kinetic} derived the Fick--Jacobs equation within the framework of mesoscopic non-equilibrium thermodynamics and proposed a heuristic scaling law for the effective diffusivity,

\begin{equation} 
D_s(x)=\frac{D_0}{\left[1+w'(x)^2\right]^{\alpha}} 
\end{equation} 

where the exponent $\alpha$ depends on the dimensionality of the confinement and was selected to recover the leading-order correction obtained by Zwanzig. This form effectively extends the range of validity of the Fick--Jacobs approximation by incorporating geometric effects beyond the first-order expansion. Subsequently, Kalinay and Percus (2006) \cite{kalinay2006corrections} developed a systematic mapping procedure that generates higher-order corrections to the effective diffusivity. In their formulation, the diffusivity is expressed as an expansion involving successive derivatives of the accessible cross section, $A'(x)$, $A''(x)$, $A'''(x)$, and higher-order terms. However, these approaches remain rooted in local derivative expansions of the confinement geometry. For obstacle-based geometries, such as channels formed by circular particles, the derivatives of the accessible area function $A(x)$ become singular near the obstacle boundaries, obscuring the physical interpretation of the resulting correction coefficients.\cite{dagdug2012diffusion}

Motivated by these limitations, we reconsider the correction from a purely geometric perspective. The curvature-corrected obstruction profile is defined as 

\begin{equation} 
\phi_{\mathrm{arc}}(r) = \phi_0\,g(r), \qquad g(r)\equiv\frac{\theta(r)}{\theta(R)}, 
\end{equation} 

where $g(r)$ represents the normalized angular obstruction profile. Introducing the normalized coordinate across the particle, 

\begin{equation} 
u=\frac{r-R}{a}, \qquad -1\le u\le 1, 
\end{equation} 

and expanding the exact circle-on-circle geometry in the limit $a/R\ll1$, one finds 
\begin{equation} 
g(u) = \frac{\theta(r)}{\theta(R)} = \sqrt{1-u^2}
\label{eq:expansion}
\end{equation} 

which corresponds to the normalized circular chord profile. The first two moments of this profile are 

\begin{equation} \langle g\rangle = \frac{1}{2} \int_{-1}^{1} \sqrt{1-u^2}\,du = \frac{\pi}{4}, \end{equation} and 
\begin{equation} \langle g^2\rangle = \frac{1}{2} \int_{-1}^{1} \left(1-u^2\right)\,du = \frac{2}{3}. 
\end{equation} 

Remarkably, these values coincide with the fitted coefficients. The motivation for examining geometric moments comes from Zwanzig's correction, where the leading correction to the Fick--Jacobs approximation appears through the second moment of the transverse equilibrium probability distribution $p(y|x)$ given in Eq.~\eqref{eq:moment}. Here, $p(y|x)$ describes the probability of finding a diffusing particle at the transverse position $y$ for a fixed longitudinal coordinate $x$, and encodes information about the local channel geometry. In the present problem, the normalized obstruction profile $g(r)$ similarly encodes the geometry variation. Therefore, we examined the moments of $g(r)$, which captured geometric information analogous to that contained in $p(y|x)$. The close agreement between the fitted coefficients and the first two moments of $g(r)$ suggests that the correction may have an underlying geometric origin rather than being purely empirical. Furthermore, expanding $g(u)$ in Eq.~\eqref{eq:expansion} shows that the leading correction is odd in $u$ and therefore vanishes upon averaging. Thus, the corresponding moments have leading corrections of order $\mathcal O((a/R)^2)$ rather than $\mathcal O(a/R)$, providing a natural explanation for the weak dependence of the fitted coefficients on particle size. This suggests that diffusion in spatially varying cross-sections may admit a simple closed-form description in terms of the geometric moments of the obstruction profile.

\section{Summary and Outlook}

We investigated diffusion across particle-laden Pickering droplet interfaces using a quasi-one-dimensional description and fully resolved two-dimensional simulations. By revisiting the geometric representation of the particle layer, we derived a curvature-corrected obstruction profile $\phi_{\mathrm{arc}}$ that accounts for finite particle size on a curved droplet interface. The resulting obstruction profile accurately reproduces the available area fraction obtained from the explicit particle geometry in the simulations. This construction provides a geometrically consistent extension of the quasi-one-dimensional model.

Comparison with particle-resolved simulations showed that the curvature-corrected quasi-one-dimensional model captures the qualitative evolution of the concentration field and incorporates the influence of particle size $r_p$ reasonably well. However, the discrepancy between the reduced and fully resolved descriptions increases systematically with interfacial packing fraction $\phi_0$. This behavior indicates that particle crowding introduces transport effects that are not captured by a purely radial description of diffusion. 

Analysis of the deviations between the models revealed that geometric effects are insufficient to capture transport at high packing fractions. We found that a simple correction based on a fitted packing fraction, $\phi_0^{\mathrm{corr}}$, aligns the quasi-one-dimensional predictions with the simulations. This correction is consistent with transport pathways involving tangential diffusion around particles, which become increasingly important as $\phi_0$ increases. Because the Fick--Jacobs equation accounts only for radial diffusion, we developed a simple moment-based correction derived from the obstruction profile motivated by the literature. The resulting model is in good agreement with fully resolved simulations for particle sizes up to approximately $15\%$ of the droplet radius considered in this study.

Our results were derived for circular particles on a circular interface. The proposed curvature-corrected obstruction profile provides a foundation for future studies exploring the role of particle shape and interfacial curvature in diffusion across particle-laden interfaces. Extending the present framework to more complex particle shapes and geometries will help determine whether the moment-based correction represents a general feature of diffusion in particle laden interfaces.

\section*{Author contributions} Rohit Khandoori: conceptualization, methodology, software, validation, formal analysis, investigation, visualization, writing – original draft. Joost de Graaf: conceptualization, methodology, formal analysis, supervision, writing – review \& editing, funding acquisition. Wim Thielemans: supervision, resources, project administration, writing – review \& editing, funding acquisition.

\section*{Conflicts of interest}
There are no conflicts to declare.

\section*{Acknowledgements}
This work was supported by the Marie Skłodowska-Curie Actions Doctoral Network BIOMAC-BP (Bio-Stabilized Multi-Phase Materials as Carriers for Biopesticides), funded by the European Union's Horizon Europe research and innovation programme under Grant Agreement No.~101168698.

%%%REFERENCES%%%
\bibliography{rsc} 
\bibliographystyle{rsc} 

\end{document}